\documentclass{osa-article}

\journal{oe}
\articletype{Research Article}

\begin{document}

\title{Photosensitive chalcogenide metasurfaces supporting bound states in the continuum}

\author{Elena Mikheeva,\authormark{1,2} Kirill Koshelev,\authormark{3,4} Duk-Yong Choi,\authormark{5} Sergey Kruk,\authormark{3} Julien Lumeau,\authormark{1} Redha Abdeddaim,\authormark{1}, \\Ivan Voznyuk\authormark{2},  
Stefan Enoch,\authormark{1} and Yuri Kivshar\authormark{3,4,*}}

\address{\authormark{1}Aix Marseille Univ, CNRS, Centrale Marseille, Institut Fresnel, F-13013 Marseille, France\\
\authormark{2}Multiwave Innovation SAS, 2 Marc Donadille, 13453 Marseille, France\\
\authormark{3}Nonlinear Physics Center, Australian National University, Canberra ACT 2601, Australia\\
\authormark{4}Department of Nanophotonics and Metamaterials, ITMO University, St. Petersburg 197101, Russia\\
\authormark{5}Laser Physics Center, Australian National University, Canberra ACT 2601, Australia}

\email{\authormark{*}yuri.kivshar@anu.edu.au} 



\begin{abstract}
We study, both theoretically and experimentally, tunable metasurfaces supporting sharp Fano-resonances inspired by optical bound states in the continuum. We explore the use of arsenic trisulfide (a photosensitive chalcogenide glass) having optical properties which can be finely tuned by light absorption at the post-fabrication stage. We select the resonant wavelength of the metasurface corresponding to the energy below the arsenic trisulfide bandgap, and experimentally control the resonance spectral position via exposure to the light of energies above the bandgap.
\end{abstract}

\section{Introduction}

All-dielectric resonant metasurfaces have emerged recently as a novel research field driven by its exceptional applications for creating low-loss nanoscale devices~\cite{kivshar_all-dielectric_2018}. They have become a viable alternative to conventional refractive optical elements as well as to diffractive optics. Conventional optical elements for light control have various limitations: they are bulky and may require a complex design for non-basic functionalities. Diffractive optical elements, in their turn, mitigate some of the limitations, and they have found applications in compensating dispersion and aberrations~\cite{turunen_diffractive_1997}, beam conversion and vortex generation~\cite{hermerschmidt_binary_2007,sabatyan_generation_2014}, medical lasers, three-dimensional imaging, and laser materials processing~\cite{bauer_geometrical_2008}. However, diffractive optical components have miniaturization limits as their thickness should be comparable to the wavelength of radiation for which they are designed. Optical metasurfaces, while being compatible with conventional micro- and nanotechnology, allow to incorporate complex control over the electromagnetic space ~\cite{kruk_functional_2017}. In contrast to diffractive optics, highly-efficient metasurfaces, being of a sub-wavelength thickness, allow overcoming the miniaturization limit.

Here we study all-dielectric metasurfaces based on resonant interactions between light and Mie-resonant dielectric particles~\cite{kruk_functional_2017}.  The tight confinement of the local electromagnetic fields and multiple interferences available in resonant dielectric nanostructures and metasurfaces can boost many optical effects and offer novel opportunities for the subwavelength control of light-matter interactions. In particular, recently emerged concept of bound states in the continuum (BICs) in nanophotonics enables a simple approach to achieve high-$Q$ resonances (or quasi-BICs) for various platforms ranging from individual dielectric nanoparticles~\cite{rybin_high-q_2017} to periodic arrangements of subwavelength resonators such as metasurfaces or chains of particles~\cite{hsu_bound_2016,koshelev_meta-optics_2018,koshelev_nonradiating_2019,yesilkoy_ultrasensitive_2019,koshelev_strong_2018}. Moreover, very recently it was revealed~\cite{koshelev_asymmetric_2018} that metasurfaces created by seemingly different lattices of dielectric meta-atoms with broken in-plane inversion symmetry can support sharp high-$Q$ resonances arising from a distortion of symmetry-protected BICs. We consider metasurfaces hosting Fano resonances \cite{fano_effects_1961,enoch_theory_2012} inspired by optical bound states in the continuum being converted into the sharp leaky dielectric slab optical resonances~\cite{koshelev_asymmetric_2018}. 

For many applications, it is beneficial to design tunable metasurfaces such as a recently reported silicon-based metasurface supporting a dynamic BIC~\cite{fan_dynamic_2019}. In the current paper, we expand the material choice by exploring a platform of chalcogenide glasses possessing {\em photosensitivity}. In conjunction with high-$Q$ resonances, the photosensitivity leads to a dramatic change of optical response of the metasurfaces. Chalcogenide glasses (ChGs) are composed of chalcogen elements such as sulfur, selenium, and tellurium. Also, additional atoms such as As, Ge, Sb, Ga, Si, and P are added. ChGs are particularly interesting due to their specific optical properties being transparent into the mid-infrared up to 20 $\mu$m, depending on their chemical composition~\cite{zakery_optical_2003}. The refractive index is high $n \sim  2\div3$, and ChGs possess also a high nonlinear refractive index n$_2$. However, one of the most striking properties is their {\it photosensitivity}. Using light exposure with a wavelength below the optical bandgap, the chemical bonds can be rearranged. This effect directly modifies the optical properties of the ChG. This very specific property has led to different kinds of applications (see, e.g., Refs.~\cite{van_popta_photoinduced_2002,shen_photosensitive_2008,bourgade_large_2019}). Layers made of ChG were shown to be excellent candidates for the production of various optical elements such as ring resonators~\cite{ma_high_2015}. It was shown recently that these layers are also good candidates for the new method of production of diffractive optical elements ~\cite{joerg_fabrication_2015}. In general, tunability of these materials can be turned into a technological advantage: refractive index variation can be recorded on uniform metasurfaces as a post-fabrication step while uniformity is advantageous for such methods of mass-production as nanoimprint \cite{briere_etchingfree_2019} or self-assembly \cite{bonod_large-scale_2015}.

Here, as the first step towards a variety of photosensitive metasurfaces, we suggest and develop a novel type of resonant metasurfaces which employ the concept of bound states in the continuum, being fabricated from chalcogenide glass As$_2$S$_3$.  We design theoretically and study experimentally chalcogenide metasurfaces supporting BIC-inspired Fano resonances, and we demonstrate a photosensitive shift of the resonances upon external light exposure.

\section{Design and numerical simulations}

\begin{figure}[t]
	\centering
	\includegraphics[width=1.0\linewidth]{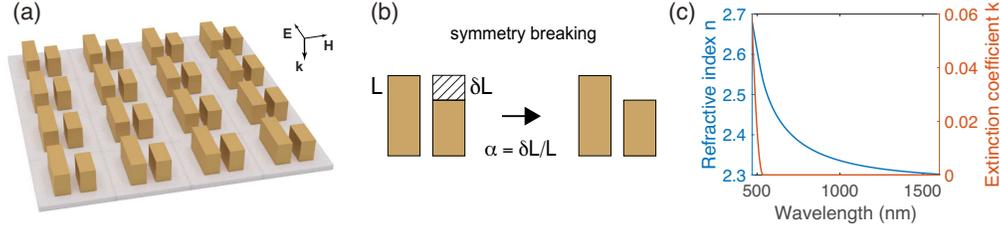}
	\caption{(a) Design of a metasurface consisting of a square array of As$_2$S$_3$-bar pairs of different length placed on a glass substrate. Parameters: the period is 490 nm, height is 180 nm, bar width is 140 nm, larger bar length is 390 nm, and the distance between bars is 100 nm. The inset shows the orientation and polarization of the incident field considered in numerical simulations and experiments. (b) Definition of the asymmetry parameter $\alpha$. (c) Dispersion of the refractive index $n$ and extinction coefficient $k$ for As$_2$S$_3$.}
	\label{fig:1}
\end{figure}

Here, we consider a metasurface made of As$_2$S$_3$ and placed on a glass substrate consisting of a square lattice of meta-atoms with broken in-plane inversion symmetry, as illustrated in Fig.~\ref{fig:1}(a). The meta-atom is constructed of a pair of rectangular bars which have lengths $L$ and $L-\delta L$, respectively. The asymmetry of the unit cell is controlled by the difference in bar lengths, which is characterized by the asymmetry parameter $\alpha = \delta L/L$, as shown in Fig.~\ref{fig:1}(b). 

First, we perform numerical analysis of linear optical spectra for the designed broken-symmetry metasurface to confirm the existence of a quasi-BIC in the optical spectral range. For transmission simulations, we employ the frequency domain solver in CST Microwave Studio and extract the transmission from the calculated S-parameters. For simulations of the eigenmode spectra, we use the eigenmode solver in COMSOL Multiphysics. All calculations are realized for a metasurface on a semi-infinite substrate surrounded by a perfectly matched layer mimicking an infinite region. The simulation area is the unit cell extended to an infinite metasurface by using the Bloch boundary conditions. All material properties for As$_2$S$_3$ are extracted from the ellipsometry data, the dispersion of the refractive index $n$ and extinction coefficient $k$ is shown in Fig.~\ref{fig:1}(c). Absorption losses are negligibly small in the region of interest and were considered to be zero. The refractive index for glass is imported from the tabulated data for SCHOTT N-BK7 glass~\cite{schott}. The incident field is a plane wave in the normal excitation geometry polarized along the long side of the bars, as shown in Fig.~\ref{fig:1}(a).

\begin{figure}[t]
	\centering
	\includegraphics[width=1.0\linewidth]{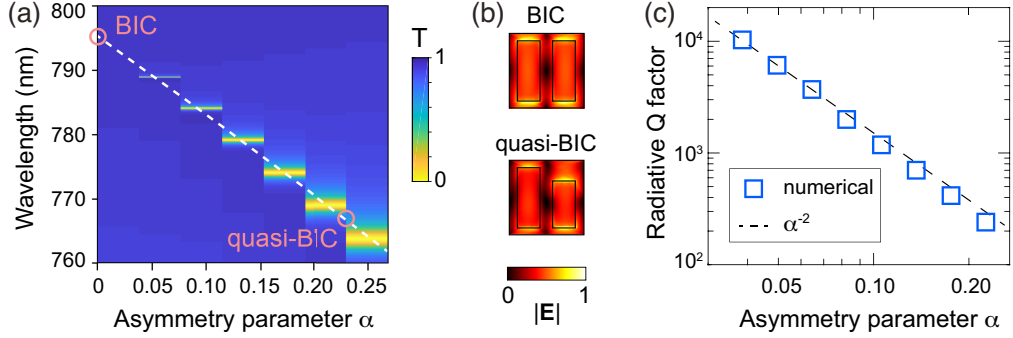}
	\caption{(a) Numerically simulated transmission spectra with respect to the excitation wavelength and $\alpha$. The white dashed line illustrates the quasi-BIC dispersion. (b) The top view of the near-field distribution of the electric field for the BIC and quasi-BIC, marked with pink circles in (a). (c) Dependence of the radiative $Q$ factor on the asymmetry parameter. The dashed line shows an inverse quadratic fitting.}
	\label{fig:2}
\end{figure}

Fig.~\ref{fig:2}(a) demonstrates the dependence of the simulated transmission spectra on the wavelength of excitation and the asymmetry parameter $\alpha$.  The white dashed line illustrates the eigenmode dispersion. The eigenmode simulations show that the metasurface with a symmetric unit cell ($\alpha=0$) supports a symmetry-protected BIC at $795$~nm, which has infinite $Q$ factor and is not manifested in the transmission spectrum. The BIC is unstable against perturbations that break the in-plane inversion symmetry, so for $\alpha>0$ it transforms into a quasi-BIC with a finite $Q$ factor~\cite{koshelev_asymmetric_2018}. The quasi-BIC is revealed in the transmission spectra as a sharp resonance with a Fano lineshape which linewidth increases with the magnitude of asymmetry. Fig.~\ref{fig:2}(b) shows the near-field distribution of the electric field for the BIC and the quasi-BIC, marked with pink circles in Fig.~\ref{fig:2}(a). The dependence of the radiative $Q$ factor on $\alpha$ follows the inverse quadratic law for small values of the asymmetry parameter~\cite{koshelev_asymmetric_2018}, as shown in Fig.~\ref{fig:2}(c). Here, we note that a quasi-BIC in a metasurface with a symmetric unit cell remains a dark mode for the far-field excitation even in the presence of absorption losses (which are considered to be zero in our simulation) or finite size of the sample. Despite fabrication imperfections allow observing specific resonant features in transmission in the vicinity of the quasi-BIC for a symmetric metasurface, such features are usually negligibly small. Hence, the meta-atom asymmetry is necessary to obtain a pronounced and sharp resonance which position and width can be adjusted by the degree of asymmetry. 

\begin{figure}[t]
	\centering
	\includegraphics[width=1.0\linewidth]{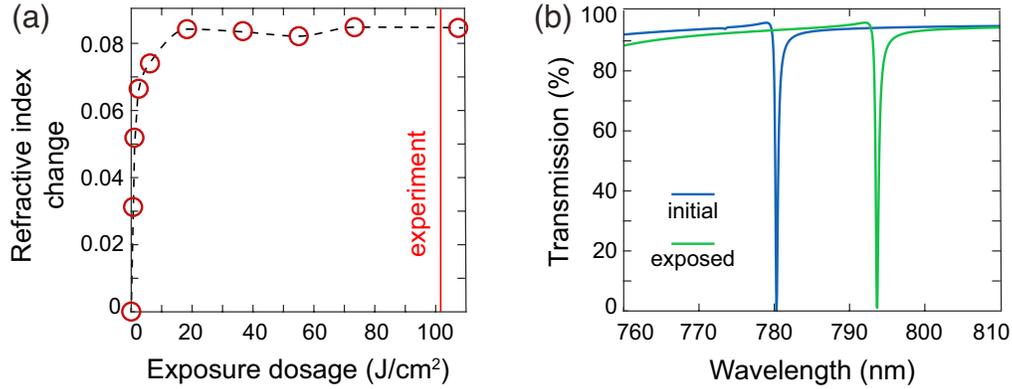}
	\caption{(a) Dependence of As$_2$S$_3$ refractive index change on the dosage of light exposure. The red open dots show the experimental results from~\cite{bourgade_large_2019}. The black dashed line is the guide for eyes. The experimental conditions are shown with the red solid line. (b) Simulated transmission spectrum for initial and exposed As$_2$S$_3$ metasurface with $\alpha=0.064$.}
	\label{fig:3}
\end{figure}

The optical properties of chalcogenide glasses can be tailored by the optical annealing via exposure to light below the material bandgap. When a sample is kept in the dark at room temperature, the induced changes are not reversible during the long period (over 1 year)~\cite{bourgade_large_2019}. However, the optical properties of arsenic trisulfide can be also controlled by the thermal annealing or by the combination of both mechanisms~\cite{choi_photo-induced_2013}. In this work, we concentrate only on the light annealing. Fig.~\ref{fig:3}(a) shows the typical dependence of As$_2$S$_3$ refractive index on the dosage of light exposure. The dependence shows a rapid growth at small exposure dosages around $10$~J/cm$^2$ and saturates for the dosages higher than $20$~J/cm$^2$. The maximal change of refractive index due to optical annealing is about $0.08$.

Next, we simulate and compare the transmission spectrum for the unexposed and exposed As$_2$S$_3$ metasurface supporting a quasi-BIC. For numerical analysis, we focus on a metasurface with a small $\alpha=0.064$ which is feasible in fabrication and provides a high-quality quasi-BIC since in the experiment we expect broadening of resonances due to the finite sample size and scattering losses.  To take into account the fabrication tolerance, we consider the difference in bar widths of $5$~nm, which follows the actual geometry of the fabricated structure, shown below. The simulated transmission spectra are shown in Fig.~\ref{fig:3}(b). The transmission curve for initial design manifests a sharp peak with a Fano lineshape at $780$~nm associated with a quasi-BIC with $Q$ factor of about $3700$. After the annealing by light with high exposure dosage, the refractive index changes by $0.08$ which leads to a $14$~nm shift of the resonant peak, as illustrated in Fig.~\ref{fig:3}(b). The $Q$ factor of the quasi-BIC supported by the annealed structure is almost unchanged. We emphasize that the predicted resonance shift is $65$ times larger than the mode linewidth because of high $Q$ factor inherent to a quasi-BIC.

\section{Fabrication and experimental results}

To demonstrate the effect of the resonant peak shift experimentally, we fabricate an As$_2$S$_3$ metasurface using the proposed design with $\alpha=0.064$. First, we deposit a $180$ nm-thick As$_2$S$_3$ thin film on a slide glass substrate (SCHOTT N-BK7)~\cite{choi_photo-induced_2013,schott}, using thermal evaporation. Next, we cover the deposited film with a $2$ nm-thick Al$_2$O$_3$ protective layer by atomic layer deposition (ALD). The pattern of bar pairs is then defined by electron-beam lithography (EBL) using positive resist (ZEP520). Prior to the resist coat, the film surface is covered with an adhesion promoter (SurPass) to improve the resist wetting and adhesion; and charge-compensating layer (E-spacer) is applied on top of the resist. After resist development, As$_2$S$_3$ film is etched in inductively coupled plasma-reactive ion etching (Plasmalab System 100, Oxford) using CHF$_3$ plasma. The residual resist is subsequently removed by an oxygen plasma. 

\begin{figure}[t]
	\centering
	\includegraphics[width=0.8\linewidth]{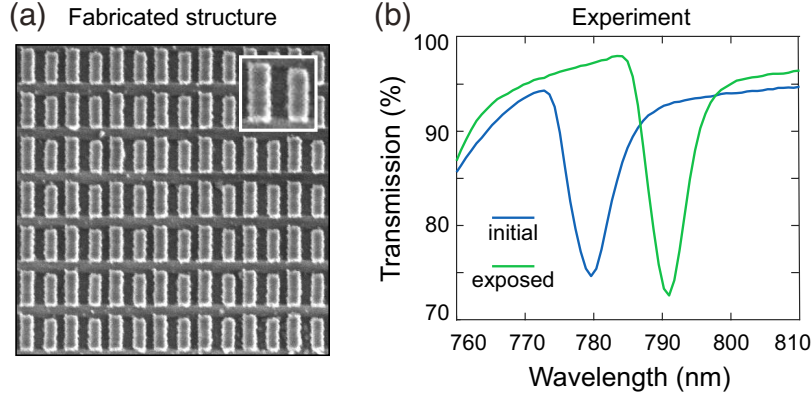}
	\caption{(a) Top-view SEM image of the As$_2$S$_3$ metasurface placed on a slide glass substrate. (b) Experimental transmission spectrum for initial and exposed As$_2$S$_3$ fabricated metasurface.}
	\label{fig:4}
\end{figure}

The fabricated metasurface structures are coated with a $2$ nm-thick Al$_2$O$_3$ film again to prevent the surface oxidation and degradation of the chalcogenide film. Figure~\ref{fig:4}(a) shows the scanning electron microscope (SEM) image of the fabricated pattern, and the inset shows the high-resolution SEM image for a single meta-atom composing the metasurface. The shorter bar width of the meta-atom of the fabricated structure is smaller than the width of the longer bar by $5$~nm.  The pattern size of each fabricated metasurface is 100~$\mu$m $\times$ 100~$\mu$m, the length of the shorter bar is $365$~nm. The fabricated sample is kept in the dark to avoid a change of the refractive index.

Next, we measure the transmission spectrum of the fabricated sample before optical annealing. For transmission measurements, we use a home-made micro-spectrometer setup with a halogen lamp as the light source. The light from the lamp passes through a 650 nm long-pass filter. The filter cuts short wavelength range preventing the annealing. The transmitted signal is analyzed with an optical spectrometer (Ocean Optics QE Pro) [see Fig. 4(b), blue curve]. The transmission spectrum is normalized to the transmission of a glass slide. After, we remove the long-pass filter and expose the metasurface with a high dosage of white light enough to achieve the saturation of the refractive index [see a vertical line in Fig.~\ref{fig:3}(a)], and measure the transmission spectrum of the exposed sample [see Fig. 4(b), a green curve].

The unexposed sample supports a quasi-BIC at $780$~nm. The resonance shifts by $12$~nm after the optical annealing. The mode positions in the experimental spectra agree well with the simulation [see Fig.~\ref{fig:3}(b)]. The $Q$ factor and peak transmittance are reduced, which may be caused by scattering losses in the fabricated sample induced by surface roughness, structural disorder, and finite size of the sample~\cite{sadrieva_transition_2017, jin_topologically_2018}. The measured $Q$ factor is extracted from the experimental transmission spectra using single-peak fitting to a Fano lineshape via the Levenberg-Marquardt algorithm. The value of the extracted $Q$ factor  is $110\pm 10$ and $120\pm 10$ for initial and exposed sample, respectively.  The error is due to the inaccuracy of the fitting procedure. The increase of the $Q$ factor for the shifted peak can be explained by the increase of the refractive index.

\section{Conclusions}

We have developed a new approach for engineering a resonant response of dielectric metasurfaces composed of meta-atoms with broken in-plane inversion symmetry closely associated with the physics of bound states in the continuum.  We have designed and studied experimentally the dielectric metasurfaces made of photosensitive chalcogenide glass As$_2$S$_3$. We have demonstrated experimentally photosensitive tunability of the metasurface resonances upon external light illumination.  The similar approach can be applied to the case of nonlinear metasurfaces~\cite{wang_nonlinear_2018,koshelev2019nonlinear} with broken-symmetry or nonlinear metasurfaces composed of arrays of chalcogenide nanoresonators designed for the nonlinear optical generation of higher harmonics. Thus, we believe that our approach is rather general, and it paves the way to smart engineering of sharp resonances in metasurfaces advanced meta-optics and nanophotonics. 
 
\section*{Funding}
Ministry of Education and Science of the Russian Federation (3.1668.2017); the Grant of the President of the Russian Federation (MK-403.2018.2); the Australian Research Council; the Strategic Fund of the Australian National University; the Foundation for the Advancement of Theoretical Physics and Mathematics BASIS;
ALPhFA: Associated Laboratory for Photonics between France and Australia.

\section*{Acknowledgments}
The authors thank Dr. Andrey Bogdanov for useful discussions and suggestions. The research was conducted within the context of the International Associated Laboratory ALPhFA: Associated Laboratory for Photonics between France and Australia. K.K. was supported by the Ministry of Education and Science of the Russian Federation (grant no. 3.1668.2017/4.6), the Grant of the President of the Russian Federation (MK-403.2018.2) and the Foundation for the Advancement of Theoretical Physics and Mathematics BASIS. Y.K. acknowledges financial support from the Australian Research Council and the Strategic Fund of the Australian National University.

\bibliography{Bibl_OE_tunable_BIC}

\begin{thebibliography}{10}
\newcommand{\enquote}[1]{``#1''}

\bibitem{kivshar_all-dielectric_2018}
Y.~Kivshar, \enquote{All-dielectric meta-optics and non-linear nanophotonics,}
  {\protect\JournalTitle{National Science Review}} \textbf{5}, 144--158 (2018).

\bibitem{turunen_diffractive_1997}
J.~Turunen and F.~Wyrowski, \emph{Diffractive {Optics} for {Industrial} and
  {Commercial} {Applications}} (John Wiley \& Sons, Limited, 1997).

\bibitem{hermerschmidt_binary_2007}
A.~Hermerschmidt, S.~Kr\"uger, and G.~Wernicke, \enquote{Binary diffractive
  beam splitters with arbitrary diffraction angles,}
  {\protect\JournalTitle{Optics Letters}} \textbf{32}, 448 (2007).

\bibitem{sabatyan_generation_2014}
A.~Sabatyan and B.~Meshginqalam, \enquote{Generation of annular beam by a novel
  class of {Fresnel} zone plate,} {\protect\JournalTitle{Applied Optics}}
  \textbf{53}, 5995 (2014).

\bibitem{bauer_geometrical_2008}
M.~Bauer, D.~Griessbach, A.~Hermerschmidt, S.~Kr\"uger, M.~Scheele, and
  A.~Schischmanow, \enquote{Geometrical camera calibration with diffractive
  optical elements,} {\protect\JournalTitle{Optics Express}} \textbf{16}, 20241
  (2008).

\bibitem{kruk_functional_2017}
S.~Kruk and Y.~Kivshar, \enquote{Functional {Meta}-{Optics} and {Nanophotonics}
  {Governed} by {Mie} {Resonances},} {\protect\JournalTitle{ACS Photonics}}
  \textbf{4}, 2638--2649 (2017).

\bibitem{rybin_high-q_2017}
M.~V. Rybin, K.~L. Koshelev, Z.~F. Sadrieva, K.~B. Samusev, A.~A. Bogdanov,
  M.~F. Limonov, and Y.~S. Kivshar, \enquote{High-{Q} supercavity modes in
  subwavelength dielectric resonators,} {\protect\JournalTitle{Physical Review
  Letters}} \textbf{119} (2017).

\bibitem{hsu_bound_2016}
C.~W. Hsu, B.~Zhen, A.~D. Stone, J.~D. Joannopoulos, and M.~Soljaci\'c,
  \enquote{Bound states in the continuum,} {\protect\JournalTitle{Nature
  Reviews Materials}} \textbf{1}, 16048 (2016).

\bibitem{koshelev_meta-optics_2018}
K.~Koshelev, A.~Bogdanov, and Y.~Kivshar, \enquote{Meta-optics and bound states
  in the continuum,} {\protect\JournalTitle{Science Bulletin}} \textbf{64},
  836--842 (2019).

\bibitem{koshelev_nonradiating_2019}
K.~Koshelev, G.~Favraud, A.~Bogdanov, Y.~Kivshar, and A.~Fratalocchi,
  \enquote{Nonradiating photonics with resonant dielectric nanostructures,}
  {\protect\JournalTitle{Nanophotonics}} \textbf{8}, 725--745 (2019).

\bibitem{yesilkoy_ultrasensitive_2019}
F.~Yesilkoy, E.~R. Arvelo, Y.~Jahani, M.~Liu, A.~Tittl, V.~Cevher, Y.~Kivshar,
  and H.~Altug, \enquote{Ultrasensitive hyperspectral imaging and biodetection
  enabled by dielectric metasurfaces,} {\protect\JournalTitle{Nature
  Photonics}} \textbf{13}, 390--396 (2019).

\bibitem{koshelev_strong_2018}
K.~L. Koshelev, S.~K. Sychev, Z.~F. Sadrieva, A.~A. Bogdanov, and I.~V. Iorsh,
  \enquote{Strong coupling between excitons in transition metal dichalcogenides
  and optical bound states in the continuum,} {\protect\JournalTitle{Physical
  Review B}} \textbf{98}, 161113 (2018).

\bibitem{koshelev_asymmetric_2018}
K.~Koshelev, S.~Lepeshov, M.~Liu, A.~Bogdanov, and Y.~Kivshar,
  \enquote{Asymmetric {Metasurfaces} with {High}- {Q} {Resonances} {Governed}
  by {Bound} {States} in the {Continuum},} {\protect\JournalTitle{Physical
  Review Letters}} \textbf{121}, 193903 (2018).

\bibitem{fano_effects_1961}
U.~Fano, \enquote{Effects of {Configuration} {Interaction} on {Intensities} and
  {Phase} {Shifts},} {\protect\JournalTitle{Physical Review}} \textbf{124},
  1866--1878 (1961).

\bibitem{enoch_theory_2012}
D.~Maystre, \enquote{Theory of {Wood's} {Anomalies},} in \emph{Plasmonics,}
  vol. 167 S.~Enoch and N.~Bonod, eds. (Springer Berlin Heidelberg, Berlin,
  Heidelberg, 2012), pp. 39--83.

\bibitem{fan_dynamic_2019}
K.~Fan, I.~V. Shadrivov, and W.~J. Padilla, \enquote{Dynamic bound states in
  the continuum,} {\protect\JournalTitle{Optica}} \textbf{6}, 169 (2019).

\bibitem{zakery_optical_2003}
A.~Zakery and S.~Elliott, \enquote{Optical properties and applications of
  chalcogenide glasses: a review,} {\protect\JournalTitle{Journal of
  Non-Crystalline Solids}} \textbf{330}, 1--12 (2003).

\bibitem{van_popta_photoinduced_2002}
A.~van Popta, R.~DeCorby, C.~Haugen, T.~Robinson, J.~McMullin, D.~Tonchev, and
  S.~Kasap, \enquote{Photoinduced refractive index change in {As}2se3 by 633nm
  illumination,} {\protect\JournalTitle{Optics Express}} \textbf{10}, 639
  (2002).

\bibitem{shen_photosensitive_2008}
W.~Shen, M.~Cathelinaud, M.~Lequime, V.~Nazabal, and X.~Liu,
  \enquote{Photosensitive post tuning of chalcogenide {Te}20as30se50 narrow
  bandpass filters,} {\protect\JournalTitle{Optics Communications}}
  \textbf{281}, 3726--3731 (2008).

\bibitem{bourgade_large_2019}
A.~Bourgade and J.~Lumeau, \enquote{Large aperture, highly uniform narrow
  bandpass {Fabry}-{Perot} filter using photosensitive {As} $_{\textrm{2}}$ {S}
  $_{\textrm{3}}$ thin films,} {\protect\JournalTitle{Optics Letters}}
  \textbf{44}, 351 (2019).

\bibitem{ma_high_2015}
P.~Ma, D.-Y. Choi, Y.~Yu, Z.~Yang, K.~Vu, T.~Nguyen, A.~Mitchell,
  B.~Luther-Davies, and S.~Madden, \enquote{High {Q} factor chalcogenide ring
  resonators for cavity-enhanced {MIR} spectroscopic sensing,}
  {\protect\JournalTitle{Optics Express}} \textbf{23}, 19969 (2015).

\bibitem{joerg_fabrication_2015}
A.~Jo\"erg and J.~Lumeau, \enquote{Fabrication of binary volumetric diffractive
  optical elements in photosensitive chalcogenide {AMTIR}-1 layers,}
  {\protect\JournalTitle{Optics Letters}} \textbf{40}, 3233 (2015).

\bibitem{briere_etchingfree_2019}
G.~Bri\`ere, P.~Ni, S.~H\'eron, S.~Chenot, S.~V\'ezian, V.~Br\"andli,
  B.~Damilano, J.~Duboz, M.~Iwanaga, and P.~Genevet, \enquote{An
  {Etching}-{Free} {Approach} {Toward} {Large}-{Scale} {Light}-{Emitting}
  {Metasurfaces},} {\protect\JournalTitle{Advanced Optical Materials}} p.
  1801271 (2019).

\bibitem{bonod_large-scale_2015}
N.~Bonod, \enquote{Large-scale dielectric metasurfaces: {Silicon} photonics,}
  {\protect\JournalTitle{Nature Materials}} \textbf{14}, 664--665 (2015).

\bibitem{schott}
\url{https://www.schott.com/}.

\bibitem{choi_photo-induced_2013}
D.-Y. Choi, A.~Wade, S.~Madden, R.~Wang, D.~Bulla, and B.~Luther-Davies,
  \enquote{Photo-induced and {Thermal} {Annealing} of {Chalcogenide} {Films}
  for {Waveguide} {Fabrication},} {\protect\JournalTitle{Physics Procedia}}
  \textbf{48}, 196--205 (2013).

\bibitem{sadrieva_transition_2017}
Z.~F. Sadrieva, I.~S. Sinev, K.~L. Koshelev, A.~Samusev, I.~V. Iorsh,
  O.~Takayama, R.~Malureanu, A.~A. Bogdanov, and A.~V. Lavrinenko,
  \enquote{Transition from {Optical} {Bound} {States} in the {Continuum} to
  {Leaky} {Resonances}: {Role} of {Substrate} and {Roughness},}
  {\protect\JournalTitle{ACS Photonics}} \textbf{4}, 723--727 (2017).

\bibitem{jin_topologically_2018}
J.~Jin, X.~Yin, L.~Ni, M.~Soljaci\'c, B.~Zhen, and C.~Peng,
  \enquote{Topologically {Enabled} {Ultra}-high-{Q} {Guided} {Resonances}
  {Robust} to {Out}-of-plane {Scattering},}
  {\protect\JournalTitle{arXiv:1812.00892}}  (2018).

\bibitem{wang_nonlinear_2018}
L.~Wang, S.~Kruk, K.~Koshelev, I.~Kravchenko, B.~Luther-Davies, and Y.~Kivshar,
  \enquote{Nonlinear {Wavefront} {Control} with {All}-{Dielectric}
  {Metasurfaces},} {\protect\JournalTitle{Nano Letters}} \textbf{18},
  3978--3984 (2018).

\bibitem{koshelev2019nonlinear}
K.~Koshelev, Y.~Tang, K.~Li, D.-Y. Choi, G.~Li, and Y.~Kivshar,
  \enquote{Nonlinear metasurfaces governed by bound states in the continuum,}
  {\protect\JournalTitle{ACS Photonics}} \textbf{6}, 1639 (2019).

\end{thebibliography}

\end{document}